\documentclass{ifacconf}
\usepackage{amssymb,amsmath,array,adjustbox,xcolor}
\usepackage{graphicx}      
\usepackage{natbib}        
\usepackage[algo2e]{algorithm2e}
\usepackage{nomencl}
\usepackage{algpseudocode}
\usepackage{algorithm}
\begin{document}
\begin{frontmatter}

\title{Attitude Control of Highly Maneuverable Aircraft Using an Improved Q-learning}


\author[First]{Mohsen. Zahmatkesh}
\author[First]{Seyyed. Ali. Emami}
\author[First]{Afshin. Banazadeh}
\author[Second]{Paolo. Castaldi}

\address[First]{Aerospace Engineering Department, Sharif University of Technology, Tehran, Iran (e-mail: banazadeh@sharif.edu).}
\address[Second]{Department of Electrical, Electronic and Information Engineering "Guglielmo Marconi", University of Bologna, Via Dell’Universit`a 50, Cesena, Italy (e-mail: paolo.castaldi@unibo.it)}

\begin{abstract}                
Attitude control of a novel regional truss-braced wing aircraft with low stability characteristics is addressed in this paper using Reinforcement Learning (RL). In recent years, RL has been increasingly employed in challenging applications, particularly, autonomous flight control. However, a significant predicament confronting discrete RL algorithms is the dimension limitation of the state-action table and difficulties in defining the elements of the RL environment. To address these issues, in this paper, a detailed mathematical model of the mentioned aircraft is first developed to shape an RL environment. Subsequently, Q-learning, the most prevalent discrete RL algorithm will be implemented in both the Markov Decision Process (MDP), and Partially Observable Markov Decision Process (POMDP) frameworks to control the longitudinal mode of the air vehicle.  In order to eliminate residual fluctuations that are a consequence of discrete action selection, and simultaneously track variable pitch angles, a Fuzzy Action Assignment (FAA) method is proposed to generate continuous control commands using the trained Q-table. Accordingly, it will be proved that by defining an accurate reward function, along with observing all crucial states (which is equivalent to satisfy the Markov Property), the performance of the introduced control system surpasses a well-tuned Proportional–Integral–Derivative (PID) controller.
\end{abstract}

\begin{keyword}
	Reinforcement Learning, Q-learning, Fuzzy Q-learning, Attitude Control, Truss-braced Wing, Flight Control
\end{keyword}

\end{frontmatter}
\section{Introduction}
The aviation industry is expeditiously growing due to world demands such as reducing fuel burn, emissions, and cost, as well as providing the faster and safer flight. This motivates the advent of new airplanes with novel configurations.
In addition, the scope clause agreement limits the number of seats in each aircraft and flight outsourcing to protect the union pilot jobs. This factor leads to an increase in production of the Modern Regional Jet (MRJ) airplane. In this regard, the importance of a safe flight becomes more vital considering more crowded airspace and new aircraft configurations having the ability to fly faster.
Truss-braced wing aircraft is one of the re-raised high-performance configurations, which has attracted significant attention from both academia \citep{li2022multipoint} and industry \citep{sarode2022investigating} due to its fuel burn efficiency.
As a result, there would be a growing need for reliable modeling and simulations, analyzing the flight handling quality, and stability analysis for such configurations \citep{nguyen2022dynamic, unknown}, while very few studies have addressed the flight control design for this aircraft.

In the last decades, various classic methods for aircraft attitude control have been developed to enhance control performance.
However, the most significant deficiency of these approaches is the insufficient capability to deal with unexpected flight conditions, while typically requiring a detailed dynamic model of the system.

Recently, the application of Reinforcement Learning (RL) has been extended to real problems, particularly, flight control design \citep{Emami.2022}.
Generally, there are two main frameworks to incorporate RL in the control design process, i.e., the high-level and low-level control systems.
In \cite{xi2022energy}, a Soft Actor-Critic (SAC) algorithm was implemented in a path planning problem for a long-endurance solar-powered UAV with energy-consuming considerations. Another work \citep{bohn2021data} concentrated on the inner loop control of a Skywalker X8 using SAC and comparing it with a PID controller. In \cite{YANG2020106100} a ANN based Q-learning horizontal trajectory tracking controller was developed based on the MDP model of an airship with fine stability characteristics. Apart from the previous method, Proximal Policy Optimization (PPO) was utilized in \cite{hu2022fixed} for orientation control of a common strongly dynamic coupled fixed-wing aircraft in the stall condition. The PPO was successful to be converged after 100000 episodes. However, useful to say that the PPO performance is adequate to optimize PID controllers \citep{dammen2022reinforcement}.

There are some papers on maneuver flight such as landing phase control both in inner and outer loops. For instance, in \cite{8695548}, a Deep Q-learning (DQL) is used to guide an aircraft to land in the desired field. In \cite{8866189}, a Deep Deterministic Policy Gradient (DDPG) was implemented for a UAV to control either path-tracking for landing glide slope and attitude control for landing flare section. Similarly, a DDPG method in \cite{9213987} is used to control outer loop of a landing procedure in presence of wind disturbance. The works which have been referred to so far accompanied ANNs to be able to converge. But to our best knowledge, there is research in attitude control using discrete RL without aiming ANNs. In \cite{10.1007/978-3-030-98404-5_59}, a Q-learning algorithm was implemented to control longitudinal and lateral angles in a general aviation aircraft(Cessna 172). This airplane profits suitable stability characteristics and also desired angles are zero. There are some fuzzy adaptations on \cite{watkins1992q} work like \cite{622790} where the Q-functions and action selection strategy are inferred from fuzzy rules. Also, \cite{1298895} proposed a dynamic fuzzy Q-learning for online and continuous tasks in mobile robots.

\begin{figure}[h!]
	\begin{center}
		\includegraphics[width=8.4cm]{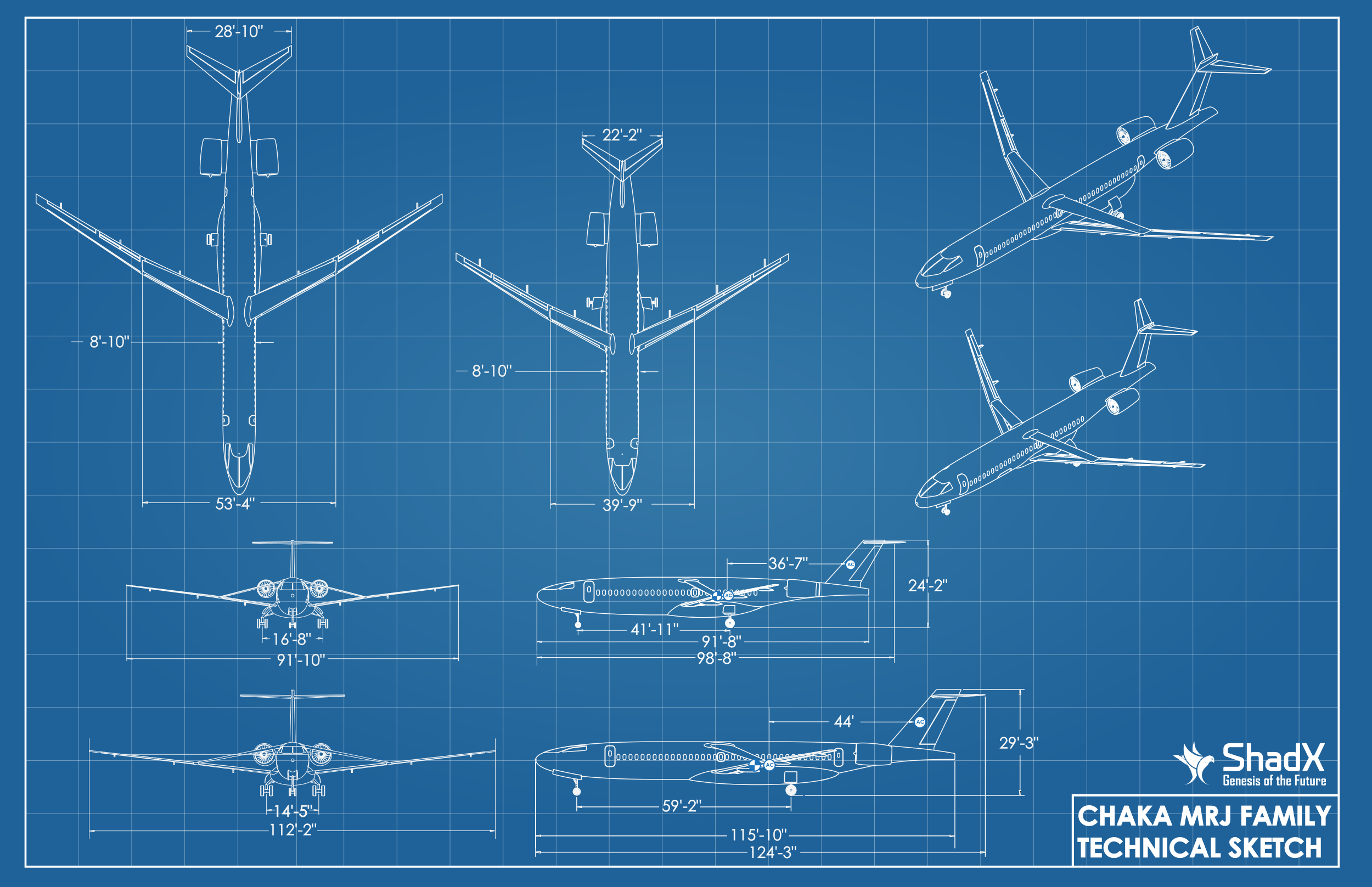}    
		\caption{Chaka MRJ Family \citep{unknown}}
		\label{fig:1}
	\end{center}
\end{figure}

Motivated by the above discussions, the main contributions of the current study can be summarized as follows: a) A truss-braced wing aircraft (Chaka 50) (\ref{fig:1}) with poor stability characteristics has been selected carefully for attitude control alongside responding to global aviation society demands. b) It will be proven that the Q-learning performance in control problems strictly depends on reward function and problem definition. So, it is able to have prosperous results even in a low stable high degree of freedom plants. c) In this work, the performance of Q-learning will be examined in both MDP and POMDP problem modelings. Also, the learned Q-table is used to generate continuous elevator deflections using Fuzzy Action Assignment (FAA) illustrating Q-table capability tracking the desired angle and also variable angles.

\section{Modeling and Simulation}
Nonlinear conservation of linear and angular momentum equations are used for modeling and simulation according to \cite{zipfel2014modeling}.
\begin{equation}\label{Eq:1}
	\begin{aligned}	
		m\begin{bmatrix}
			\dot{u} \\
			\dot{v} \\
			\dot{w}
		\end{bmatrix}+m\begin{bmatrix}
			0 & -r & q \\
			r & 0 & -p \\
			-q & p & 0
		\end{bmatrix}\begin{bmatrix}
			u \\
			v \\
			w
		\end{bmatrix}=\begin{bmatrix}
			F_{A_x+T_x} \\
			F_{A_y+T_y} \\
			F_{A_z+T_z}
		\end{bmatrix}
		+ m\begin{bmatrix}
			g_x \\
			g_y \\
			g_z
		\end{bmatrix}
	\end{aligned}
\end{equation}
\begin{equation}\label{Eq:2}
	\begin{aligned}	
		\begin{bmatrix}
			I_x & 0 & 0 \\
			0 & I_y & 0 \\
			0 & 0 & I_z
		\end{bmatrix}\begin{bmatrix}
			\dot{p} \\
			\dot{q} \\
			\dot{r}
		\end{bmatrix}+\begin{bmatrix}
			0 & -r & q \\
			r & 0 & -p \\
			-q & p & 0
		\end{bmatrix}\begin{bmatrix}
			I_x & 0 & 0 \\
			0 & I_y & 0 \\
			0 & 0 & I_z
		\end{bmatrix}\begin{bmatrix}
			p \\
			q \\
			r
		\end{bmatrix}
		=
		\begin{bmatrix}
			L_{A+T} \\
			M_{A+T} \\
			N_{A+T}
		\end{bmatrix}
	\end{aligned}
\end{equation}
where assuming the moments of thrust are equal to zero and also the thrust is only implying in $x$ direction. Therefore, $L_T = M_T = N_T = F_{T_y} = F_{T_z} = 0$. and the aerodynamic forces and moments in body axis are as follows:
\begin{equation}\label{Eq:3}
	\begin{aligned}
		\begin{bmatrix}
			F_{A_x} \\
			F_{A_z} \\
			M_A
		\end{bmatrix}^B =\bar{q} S \bar{c}
		\begin{bmatrix}
			c_{L_0} & c_{L_\alpha} & c_{L_{\dot{\alpha}}} & c_{L_u} & c_{L_q} & c_{L_{\delta_E}}\\
			c_{D_0} & c_{D_\alpha} & c_{D_{\dot{\alpha}}} & c_{D_u} & c_{D_q} & c_{D_{\delta_E}}\\
			c_{m_0} & c_{m_\alpha} & c_{m_{\dot{\alpha}}} & c_{m_u} & c_{m_q} & c_{m_{\delta_E}}
		\end{bmatrix}
		\begin{bmatrix}
			1\\
			\alpha\\
			\frac{\dot{\alpha} \bar{c}}{2 V_{P_1}}\\
			\frac{u}{V_{P_1}}\\
			\frac{q \bar{c}}{2 V_{P_1}}\\
			\delta_E
		\end{bmatrix}
	\end{aligned}
\end{equation}
The vector of gravity acceleration in the body axis defined by (\ref{Eq:1}) is as follows:
\begin{equation}\label{Eq:4}
	\begin{bmatrix}
		g_x\\
		g_y\\
		g_z
	\end{bmatrix}^B =
	\begin{Bmatrix}
		-g\sin(\theta)\\
		g\cos(\theta)\sin(\phi)\\
		g\cos(\theta)\cos(\phi)
	\end{Bmatrix}
\end{equation}
And also, the rotational kinematic equations are necessary for transfer from body to inertia coordinates.
\begin{equation}\label{Eq:5}
	\begin{bmatrix}
		\dot{\phi} \\ \dot{\theta} \\ \dot{\psi}
	\end{bmatrix}=\begin{bmatrix} 1 & \sin \varphi \tan \theta & \cos \varphi \tan \theta \\ 0 & \cos \varphi & -\sin \varphi \\ 0 & \sin \varphi / \cos \theta & \cos \varphi / \cos \theta \end{bmatrix}\begin{bmatrix} p \\ q \\ r
	\end{bmatrix}
\end{equation}
using (\ref{Eq:1}), and (\ref{Eq:5}), velocity vector in inertia coordinate is achievable. For contraction reasons; $\sin = \textup{s}, \cos = \textup{c}$.
\begin{equation}\label{Eq:6}
	\begin{bmatrix} \dot{x} \\ \dot{y} \\ \dot{z} \end{bmatrix}^I=\begin{bmatrix} \textup{c} \psi \textup{c} \theta & \textup{c} \psi \textup{s} \theta \textup{s} \varphi-\textup{s} \psi \textup{c} \varphi & \textup{c} \psi \textup{s} \theta \textup{c} \varphi+\textup{s} \psi \textup{s} \varphi \\ \textup{s} \psi \textup{c} \theta & \textup{s} \psi \textup{s} \theta \textup{s} \varphi+\textup{c} \psi \textup{c} \varphi & \textup{s} \psi \textup{s} \theta \textup{c} \varphi-\textup{c} \psi \textup{s} \varphi \\ -\textup{s} \theta & \textup{c} \theta \textup{s} \varphi & \textup{c} \theta \textup{c} \varphi \end{bmatrix}\begin{bmatrix} u \\ v \\ w \end{bmatrix}^B
\end{equation}
Stability and control derivatives for the Chaka-50 are reported in \cite{unknown} based on Computational Fluid Dynamics (CFD). The summary of these derivatives for two flight phases is given in table (\ref{tab:table1}). before six-degree-of-freedom (6DoF) simulation using equations (\ref{Eq:1}-\ref{Eq:2}), the trim conditions in a wings-level flight are calculated for simulation verification based on trim equations in \cite{roskam1998airplane}. In drag equation, absolute value of $\delta_E$, $i_{H_1}$, $\alpha_1$  is considered. Also, flight path angle $\gamma_1$,  motor installation angle $\phi_T$, and horizontal tail incidence angle $i_H$, are zero.
The elevator deflection $\delta_E$, and required thrust $T_1$ for a trim flight is obtained and shown in table (\ref{tab:table2}).
The values in the table (\ref{tab:table2}) are important for 6DoF simulation validation.
\subsubsection{Atmospheric Disturbance and Sensor Measurement Noise}
This research has utilized the Dryden atmosphere turbulence for its simple mathematical modeling.
\begin{equation}\label{Eq:7}
    G_w(s) = \sigma_w\sqrt{\frac{L_w}{\pi u_1}}\Bigg[\frac{1+\sqrt{3}\frac{L_w}{u_1}s}{1+(\frac{L_w}{u_1}s)^2}\Bigg]
\end{equation}
This model is applied in $w$ direction where $L_w=h=100m$, $\sigma_w=10$, and $u_1=160 \frac{m}{s}$. In addition, the sensor noise is defined as $10\%$ of sensor measurement.
\begin{table}[h!]
	\begin{center}
		\caption{Stability, Control derivatives (1/rad)}
		\label{tab:table1}
		\begin{tabular}{|c|c|c|} 
			\hline
			\begin{tabular}{c}Longitudinal\\Derivatives
			\end{tabular} & Take-off & Cruise\\
			\hline
			$c_{D_0}$ & 0.0378 & 0.0338  \\
			$c_{L_0}$  & 0.3203 & 0.3180\\
			$c_{m_0}$ & -0.07 & -0.06  \\
			$c_{D_\alpha}$ & 0.95 & 0.8930  \\
			$c_{L_\alpha}$ & 11.06 & 14.88 \\
			$c_{m_\alpha}$ & -12.18 & -11.84  \\
			$c_{D_u}$ & 0.040 & 0.041  \\
			$c_{L_u}$ & 0 & 0.081\\
			$c_{m_u}$ & 0 & -0.039\\
			$c_{D_q}$ & 0 & 0\\
			$c_{L_q}$ & 11.31 & 12.53  \\
			$c_{m_q}$ & -40.25 & -40.69  \\
			$c_{D_{\delta_E}}$ & 0.1550 & 0.1570 \\
			$c_{L_{\delta_E}}$ & 0.96 & 0.78  \\
			$c_{m_{\delta_E}}$ & -6.15 & -5.98 \\
			\hline
		\end{tabular}
	\end{center}
\end{table}
\begin{table}[h!]
	\begin{center}
		\caption{Trim Conditions of Chaka MRJ}
		\label{tab:table2}
		\begin{tabular}{|c|c|c|} 
			\hline
			\begin{tabular}{c} Required Thrust\\ ($T_1$)(lbs) \end{tabular}  & AoA ($\alpha^\circ$) & \begin{tabular}{c}Required Elevator \\($\delta_E^\circ$) \end{tabular} \\
			\hline
			21433.02 & 0.39 & -2.28 \\
			\hline
		\end{tabular}
	\end{center}
\end{table}
Also, the geometric, mass, and moment of inertia data are given in the table (\ref{tab:table3}).
\begin{table}[h!]
	\begin{center}
		\caption{Simulation Parameters}
		\label{tab:table3}
		\begin{tabular}{|c|c|c|c|} 
			\hline
			Parameter  & Value & Parameter & Value \\
			\hline
			Wing Area($m^2$) & 43.42 & $I_{xx}$($kg.m^2$) & 378056.535 \\
			\hline
			\begin{tabular}{c}
				Mean Aerodynamic \\ Chord($m$)
			\end{tabular} & 1.216 & $I_{yy}$($kg.m^2$) & 4914073.496 \\
			\hline
			Span($m$) & 28 & $I_{zz}$($kg.m^2$) & 5670084.803 \\
			\hline
			Mass($kg$) & 18418.27 & $I_{xz}$($kg.m^2$) & 0 \\
			\hline
		\end{tabular}
	\end{center}
\end{table}
\section{Attitude Control Process Using Q-learning}
Q-learning is an off-policy, model-free control strategy algorithm that by interacting with an environment, seeks to find the best action to take given the current state. It is a branch of Temporal Difference (TD) which is a combination of Dynamic Programming, and Monte Carlo theories.
Truss-braced wing aircraft usually suffer inadequate stability owing to their narrow mean aerodynamic chord (MAC). For example, the Phugoid and Short Period poles of Boeing N+3 TTBW \citep{{nguyen2022dynamic}}, and Chaka 50 \citep{{unknown}} prove that intuitively in comparison with corresponding poles in \cite{cetin2018} for the Cessna 172. A summary of numerical results has been gathered in table (\ref{tab:table4}).
\begin{table}[h!]
	\begin{center}
		\caption{Longitudinal Dynamics Characteristic}
		\label{tab:table4}
		\begin{tabular}{|c|c|c|} 
			\hline
			Aircraft\ Roots& Short Period Roots & Phugoid Roots \\
			\hline
			Chaka 50 & $-0.8 \pm 0.61i$ & $-0.0064 \pm 0.05i$ \\
			\hline
			Cessna 172 & $-3.23 \pm 5.71i$ & $-0.025 \pm 0.19i$  \\
			\hline
			Boeing N+3 & $-0.35 \pm 0.35i$ & $-0.0082 \pm 0.07i$  \\
			\hline
		\end{tabular}
	\end{center}
\end{table}
\subsection{MDP and POMDP Definition in Attitude Control}
It is necessary to formalize sequential decision making like aircraft attitude control as MDPs where one action influence not just next state and its immediate reward, but also upcoming states and their future rewards \citep{{sutton2018reinforcement}}. For clarification, At each time-step $t$, the controller normally receives the state's information including $\theta_t \in \mathcal{S}_1$, and $\dot{\theta_t} \in \mathcal{S}_2$ from the environment. Based on that, the controller selects an action which in this model is the elevator deflection, $\delta_{E_t} \in \mathcal{A}(s)$. The simulation executes and in next time-step $t+1$, the controller receives a numerical reward, $R_{t+1} \in \mathcal{R}$ to evaluate its performance and find itself in next state, $\theta_{t+1}$, $\dot{\theta}_{t+1}$.
\begin{equation}\label{Eq:8}
	\theta_0,\ \dot{\theta}_0,\ \delta_{E_0},\ R_1,\ \theta_1,\ \dot{\theta}_1,\ ...
\end{equation}
In this problem, a random selection of $R_t$, $\theta_t$, and $\dot{\theta}_t$ have a clear discrete probability distribution dependent on previous state and action only. Also, the problem has Markov property by considering $\theta_t$, and $\dot{\theta}_t$ as states.
The purpose of reinforcement learning finite MDP is to find a policy that gathers maximum reward over time. Consequently, to find an optimal policy for taking $\delta_E$ in state $\theta$, $\dot\theta$, the state-action value function $Q_\pi(\theta, \dot\theta,a)$ defines to be maximize by expected return that is sum of discounted instant rewards by starting from one specific state following policy $\pi$ to terminal state $\theta_T$, $\dot{\theta}_T$.
\begin{equation}\label{Eq:9}
	\begin{aligned}
		Q_\pi(\theta,\,\dot\theta,\,\delta_E) =  \mathbb{E}_\pi \bigg[ \sum_{k=0}^{\infty}\gamma^k & R_{t+k+1}\ \bigg|\  \\ & \theta_{t} = \theta,\, \dot{\theta}_t = \dot{\theta},\, \delta_{E_t}=\delta_E \bigg],
	\end{aligned}
\end{equation}
Where $\gamma$ is the discount factor for rewards to be weighted based on its time-step, and usually is $0<\gamma<1$.
\subsection{Structure of Q-learning Controller}
In this work, the optimal policy of elevator action selection in each state is approximated directly using an early breakthrough in reinforcement learning namely Q-learning \citep{watkins1992q}. Because of the non-linearity and poor stability characteristics of Truss-braced wing aircraft, Q-learning implementation without utilizing NNs can be challenging.

\subsubsection{Reward Function and Action Space Definition}

Defining an efficient reward function plays a main role in algorithm convergence. Therefore, this research has concentrated carefully in reward function design and hyper parameter tuning. In this way, the reward function contains different components such as $\theta,\, q,\, \delta_E$.
The reward function would be computed in three consecutive steps.
First, to restrict the operating frequency of the elevator, it is essential to give a large punishment in the case of aggressive elevator selection:
\begin{equation}\label{Eq:11}
Reward_t =-10000, \quad \textrm{If} \ \left(|\delta_{E_t}|-|\delta_{E_{t-1}}|\right)>0.1\ \textrm{rad}.
\end{equation}

Subsequently, if the change rate of the elevator is satisfactory, the reward function will be computed as follows if the aircraft is in the vicinity of the desired state.
\begin{equation}\label{Eq:12}
	\begin{aligned}
		Reward_t =\
        & (300, \qquad \textrm{If} \ |err_{p_t}|< 0.05^\circ)   \\ + \
        & (300, \qquad \textrm{If} \  |err_{p_t}|< 0.02^\circ) \ \\+ \
		& (400, \qquad \textrm{If} \ |q_{sim_t}|< 0.04^\circ) \ \\+ \
		& (600, \qquad \textrm{If} \ |q_{sim_t}|< 0.02^\circ) \ \\+ \
		& (800, \qquad \textrm{If} \ |q_{sim_t}|< 0.005^\circ), \
	\end{aligned}
\end{equation}
where $err_{p_t} = \theta_{sim_t} - \theta_{des_t}$.
Finally, if none of the above two conditions are met, we should encourage the air vehicle to move towards the desired state. This can be done using the following reward function:
\begin{equation}\label{Eq:13}
	Reward_t = - (100 \times |err_{p_t}|)^2 - (40 \times |q_{sim_t}|)^2.
\end{equation}

Accordingly, the farther the system is from the desired state, the less reward it receives. Also, a derivative term (the second term) has been incorporated into the reward function to avoid high pitch rates.

Considering the tabular Q-learning, the elevator commands are obtained discretely. So, the elevator commands are divided into $-0.25$ to $+0.25$ radians with $0.025$ intervals, corresponding 21 elevator deflections. Also, $\epsilon$-greedy action selection strategy with epsilon decay is used in this research.
\begin{equation}\label{Eq:14}
	\delta_{E_t} =
	\begin{cases}
		\arg\max \, Q(\theta_{t},\, \dot\theta_{t},\, \delta_{E}) & \text{with probability $1-\epsilon$}\\
		\text{random action} & \text{with probability $\epsilon$}\\
	\end{cases}
\end{equation}
The mentioned reward function has developed carefully during lots of dynamic examinations. Against some works, this research believes that the quality of learning convergence not only is not related to trial and error but also, is related to deep comprehension of dynamic feedback. For clarification, we first were omitting to observe $q_{sim}$ feedback in equation (\ref{Eq:12}) that caused worse results. Also, the proportion of the first term to the second in equation (\ref{Eq:13}) plays a substantial role in the convergence rate.
\subsection{Structure of Fuzzy Action Assignment}

In order to make continuous elevator actions, a Fuzzy Action Assignment (FAA) is presented to evaluate the capability of the learned Q-table. In this method, instead of taking a discrete greedy action in a given state $\theta$, $\dot{\theta}$, a weighted $\delta_{E}$ is selected based on the proposed membership function (MF). More precisely, the membership function corresponding to a cell of the table with $\theta_i$ and $\dot{\theta}_j$ is defined as follows:
\begin{equation}\label{Eq:15}
	\begin{aligned}
		& MF_{i,j} = \\	& \exp\left(-\frac{1}{2}\left(\frac{\theta_{sim_{t}}-\theta_i}{\sigma_{\theta}}\right)^2\right)  \exp\left(-\frac{1}{2}\left(\frac{q_{sim_{t}}-\dot{\theta}_j}{\sigma_{\dot{\theta}}}\right)^2\right),
	\end{aligned}
\end{equation}
Where $\sigma_{\theta_{i}} = \frac{\theta_i - \theta_{i-1}}{2}$, and $\dot{\sigma}_{\theta{j}} = \frac{\dot{\theta}_j - \dot{\theta}_{j-1}}{2}$  are the half-length of the state span, provided that $\theta_i > \theta_{i-1}$, and $\dot{\theta}_j > \dot{\theta}_{j-1}$. Also, $\theta_{n} = \frac{\theta_i + \theta_{i-1}}{2}$, and $\dot{\theta}_{n} = \frac{\dot{\theta}_j + \dot{\theta}_{j-1}}{2}$ are the center of each state. So the $\delta_{E}$ at each time-step is calculated as:
\begin{equation}\label{Eq:16}
	\begin{aligned}
		\delta_{E_t}=\frac{\sum_i\sum_j MF_{i,j} \arg\max \, Q(\theta_{i},\, \dot\theta_{j},\, \delta_{E})}{\sum_i\sum_j MF_{i,j}}.
	\end{aligned}
\end{equation}

\section{Simulation Results and Discussion}

\begin{figure}
	\begin{center}
		\includegraphics[width=9.0cm]{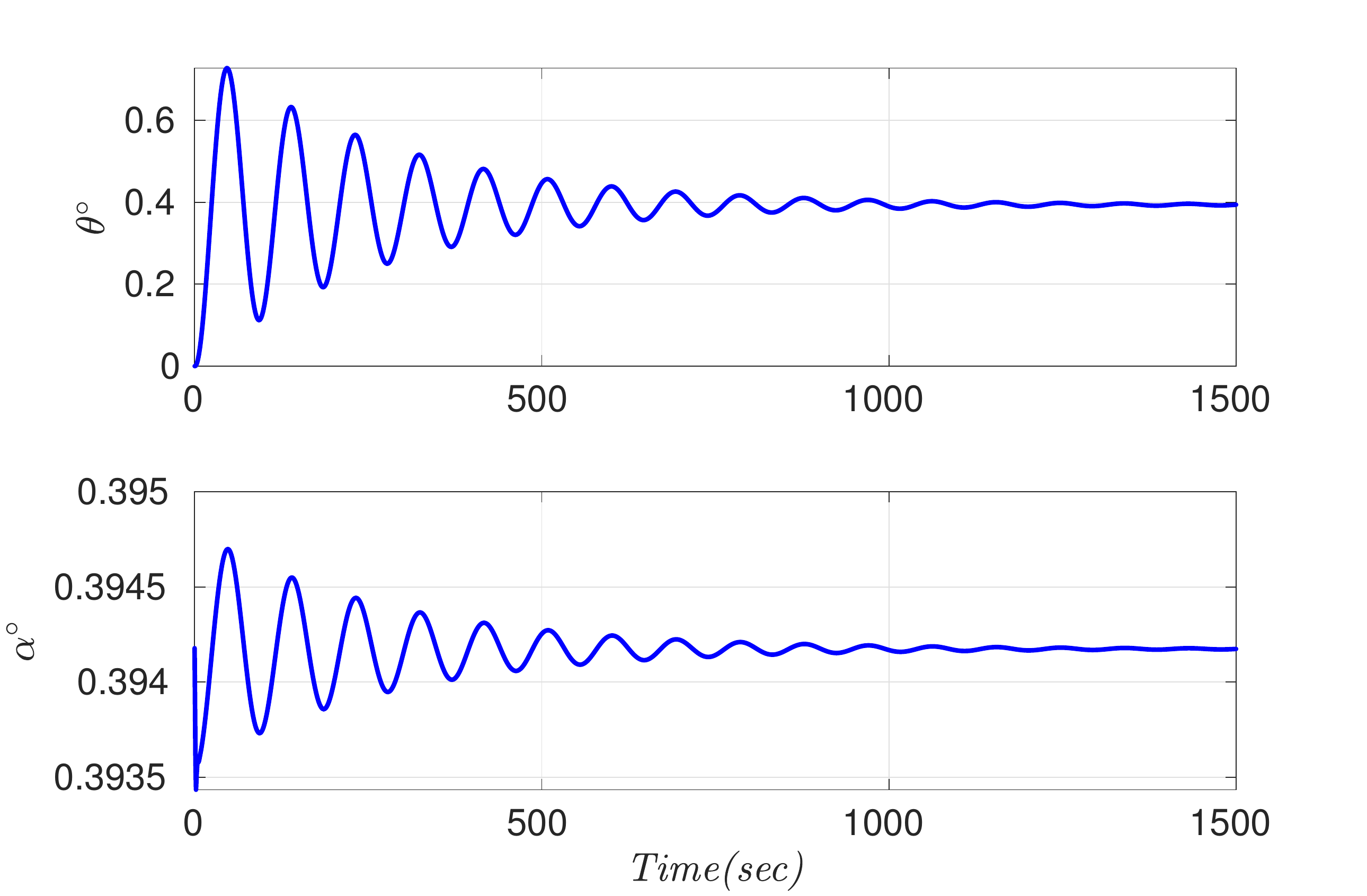}    
		\caption{The $\alpha$ and $\theta$ variation in a trim flight}
		\label{fig:2}
	\end{center}
\end{figure}

The attitude control problem is divided into two parts. First, the environment includes the aircraft simulation, and second, the Q-learning agent which is a controller. In general, at each time step, the $\theta_{sim}$ is obtained, and then the calculated tracking error is used to rectify the action selection policy. It is continued for several episodes so as to reach an optimal strategy. The pseudocode of the presented method is as follows:
\begin{algorithm}
    \caption{Q-learning Attitude Controller}
    \SetAlgoLined
    \DontPrintSemicolon
    \textbf{Input:} Learning Rate $\alpha$, Discount Factor $\gamma$, Desired Angle $\theta_{des}=1$deg.\newline
    \textbf{Output:}\ $Q_{\pi^*}(\theta, \dot\theta, \delta_{E})$ \newline
    //Initialize\ $Q(\theta_0, \dot\theta_{0}, \delta_{E_0})\gets\emptyset,\ $for all$\ \theta\in\mathcal{S}_1,\dot{\theta}\in\mathcal{S}_2, \delta_E\in\mathcal{A}(s)$

    \For  {each episode (5 $sec$)} {
        // Initialize 6 DoF simulation with a random $\theta_0 \in [0,2]$ deg.\newline
        \For{each time-step (0.01 $sec$)}{(1) Select an action $\delta_E$ based on FAA $\epsilon$-greedy strategy.\newline
        (2) Execute 6 DoF simulation using computed  $\delta_E$, observe $R_{t+1}$, $\theta_{t+1}$, $\dot\theta_{t+1}$.\newline
        (3) Update the state-action value function:
        \begin{align*}& Q(\theta_{t},\, \dot{\theta_t}, \, \delta_{E_t}) =  \, Q(\theta_{t},\, \dot{\theta_t}, \, \delta_{E_t}) +\, \\& \alpha \bigg[R_{t+1}+ \gamma \, \underset{\delta_{E}}{max} \, Q(\theta_{t+1},\, \dot\theta_{t+1}, \, \delta_{E})-Q(\theta_{t},\, \dot{\theta_t}, \, \delta_{E_t})\bigg]
        \end{align*}\newline
        (4) Substitute simulation parameters in time-step $t$ with $t+1$.}
    }
    return $Q_{\pi^*}(\theta, \dot\theta, \delta_{E})$
    \label{alg:PoEG}
\end{algorithm}
\newline
The above method is applied to Chaka 50 MRJ in 5 second episodes. The difference between the proposed control theory and trim flight proves the performance of the Q-learning controller. Figure (\ref{fig:2}) illustrates the simulation results of trim conditions over 1500 seconds. Needless to say, the $\alpha$ angle is converged exactly to its theoretical value by the table (\ref{tab:table2}). Also, the $\theta$ angle is converged in accordance with $\alpha$. Consequently, the simulation is accurate. But low-stability existence resulting in long-time fluctuations is affected by the damping ratio of the Phugoid mode.

\begin{figure}
	\includegraphics[width=9.0cm]{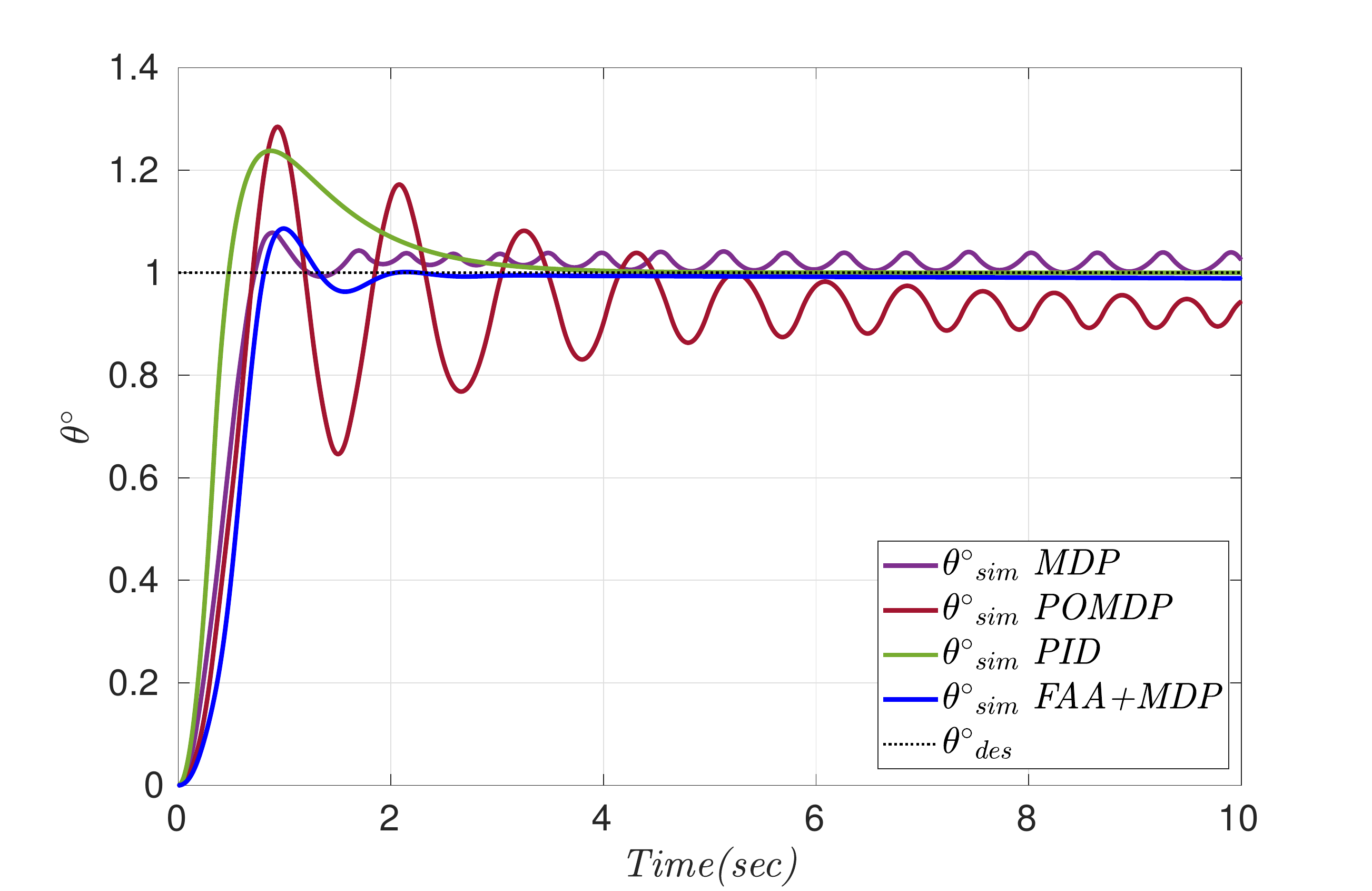}
	\caption{Performance of improved MDP Q-learning using FAA, in comparison to MDP, POMDP, and PID controller.}
	\label{fig:3}
\end{figure}

\begin{figure}
	\includegraphics[width=9.0cm]{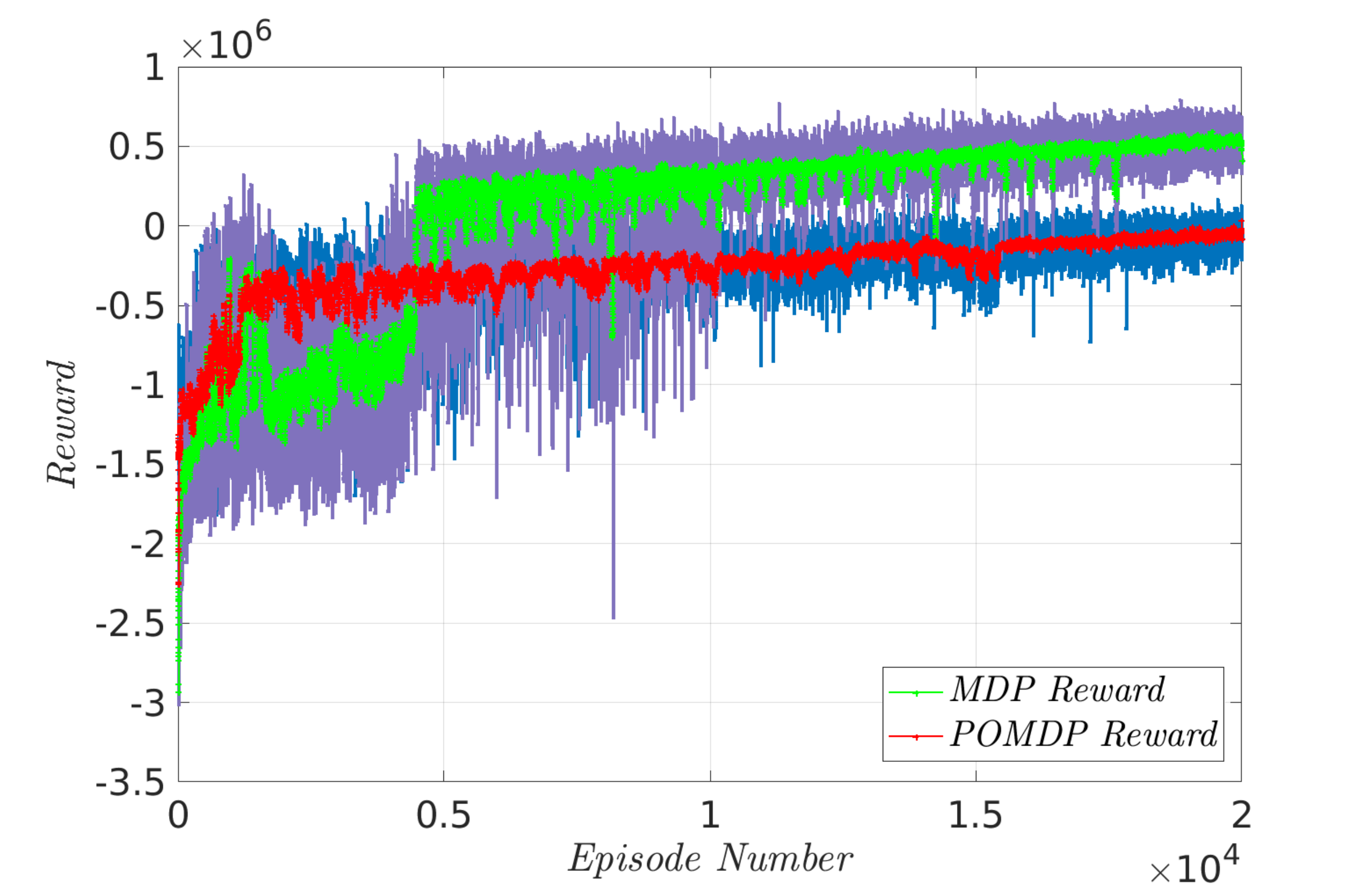}
	\caption{MDP and POMDP rewards over $20000$ episodes }
	\label{fig:5}
\end{figure}

In this way, the simulations using the controller are performed in two problem modelings as MDP and POMDP. The difference between them is observing $\dot{\theta}$ in the MDP model where the POMDP is neglected. Obviously, the state-action table in the MDP model is 3D as $\theta \times \dot{\theta} \times \delta_{E}$ dimension whereas in the POMDP model, this table is shaped as dimension as $\theta \times \delta_{E}$. In addition to the number of observed states, the bounds, and intervals of states are important in converging time. In this way, to learn the RL controller efficiently, it is momentous to divide $\theta$, $\dot{\theta}$ intervals knowledgeably so as not to omit any main state. As a synopsis, the simulation parameters and the intervals are in table (\ref{Eq:5})
\begin{table}
	\begin{center}
		\caption{Q-learning Simulation Parameters}
		\label{tab:table5}
		\begin{tabular}{|c|c|}
			\hline
			\textbf{Parameter} & \textbf{Value}  \\
			\hline
			Epsilon($\epsilon$) & $[0.1: \textcolor{blue}{3e{-6}}: 0.04]$  \\
			\hline
			Alpha($\alpha$) & $[0.02: \textcolor{blue}{9e{-7}}: 0.002]$ \\
			\hline
			Gamma($\gamma$)& $0.99$ \\
			\hline
			Episode number& $20000$ \\
			\hline
			$\theta$(rad) & $[-10, -0.024: \textcolor{blue}{0.002}: -0.002, -0.001, 0]$\\
			\hline\\[-1.1em]
			$\dot\theta$(rad) & $[-10, -0.04, -0.02, -0.005]$\\
			\hline
		\end{tabular}
	\end{center}
\end{table}
where the $\epsilon$, and $\alpha$ are quantified with their upper bounds in the first episode, and decay at a linear rate during episodes. Also, the state-action table intervals including $\theta$, and $\dot{\theta}$ are mentioned in this table where blue numbers are intervals. These values are considered symmetrically with positive signs for the positive zone. Assuming this, all substantial states that the aircraft confronts during the learning phase are covered. The result of the Q-learning controller is illustrated in figure (\ref{fig:3}) in comparison to the PID controller. Obviously, the POMDP accuracy is worse than others because the environment modeling is lacking from a complete Markov Property. However, the POMDP performance is significantly better than the trim condition in figure (\ref{fig:2}). It is apparent that PID rise-time is lower than MDP and POMDP but the overshoot of PID continues to exist until approximately 2 seconds. There is a tiny oscillation in the MDP method because of discrete elevator deflections, but this flaw is eliminated by FAA using the same learned Q-table.
Figure (\ref{fig:5}) shows the rewards of each episode for MDP and POMDP modeling. In early episodes, POMDP results are better and making fewer fluctuations. But after about 4000 episodes, MDP starts achieving positive rewards. The MDP converges fairly in episode number 10000 and surpasses the POMDP method. It is noticeable that the POMDP never achieves positive rewards. Consequently, encompassing the $\dot{\theta}$ plays a significant role in efficiency.
\begin{figure}
	\includegraphics[width=9.0cm]{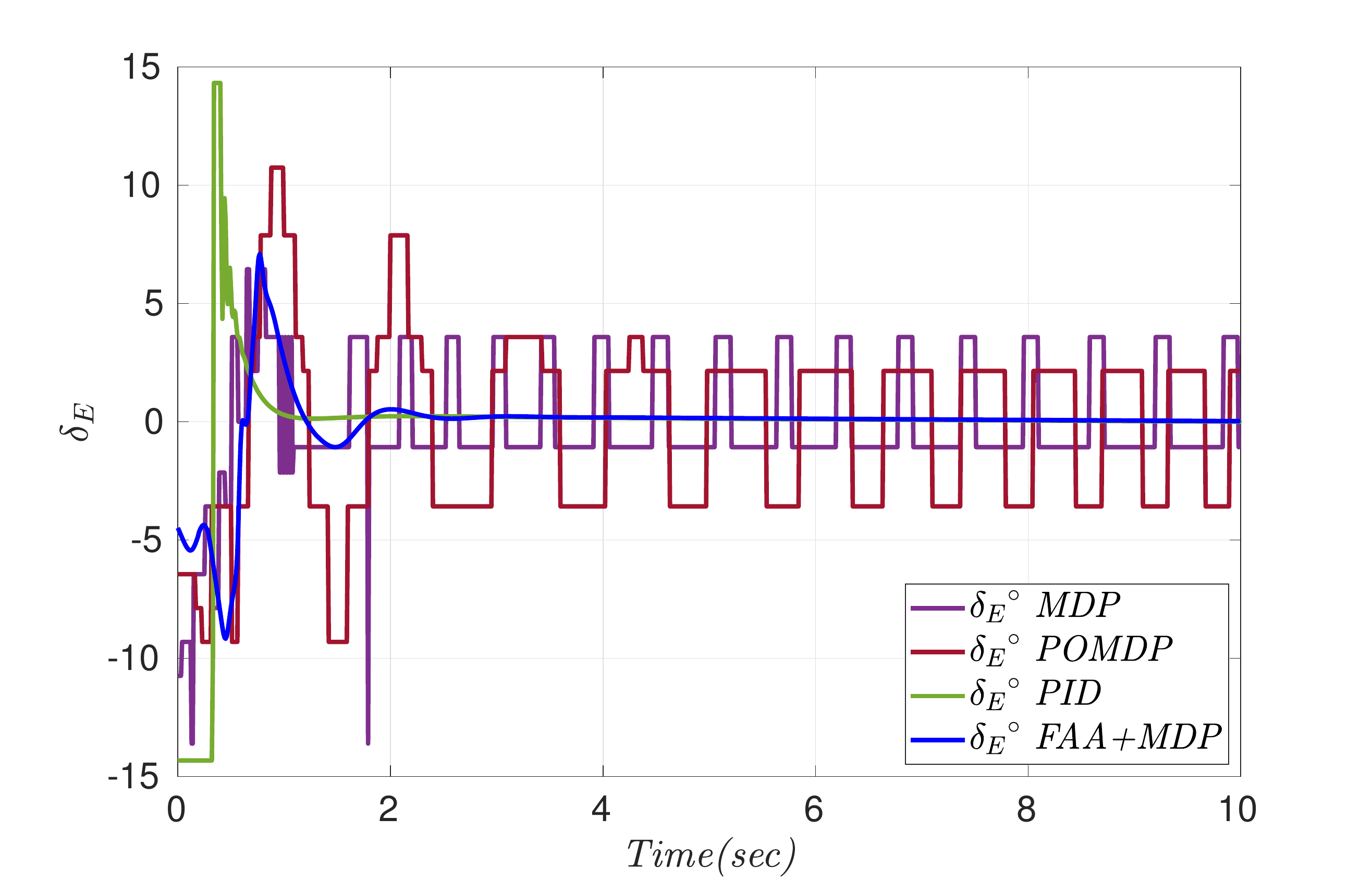}
	\caption{Elevator deflections in $\theta_{des} = 1^\circ$ tracking}
	\label{fig:6}
\end{figure}
Elevator deflections is illustrated in figure (\ref{fig:6}). Although the control effort of MDP is more than POMDP numerically, it sequels better $\theta_{does}$ tracking results. However, the FAA solves the consequent oscillations caused by discrete actions simultaneously reducing control effort even more superior to PID.
\begin{figure}
	\includegraphics[width=9.0cm]{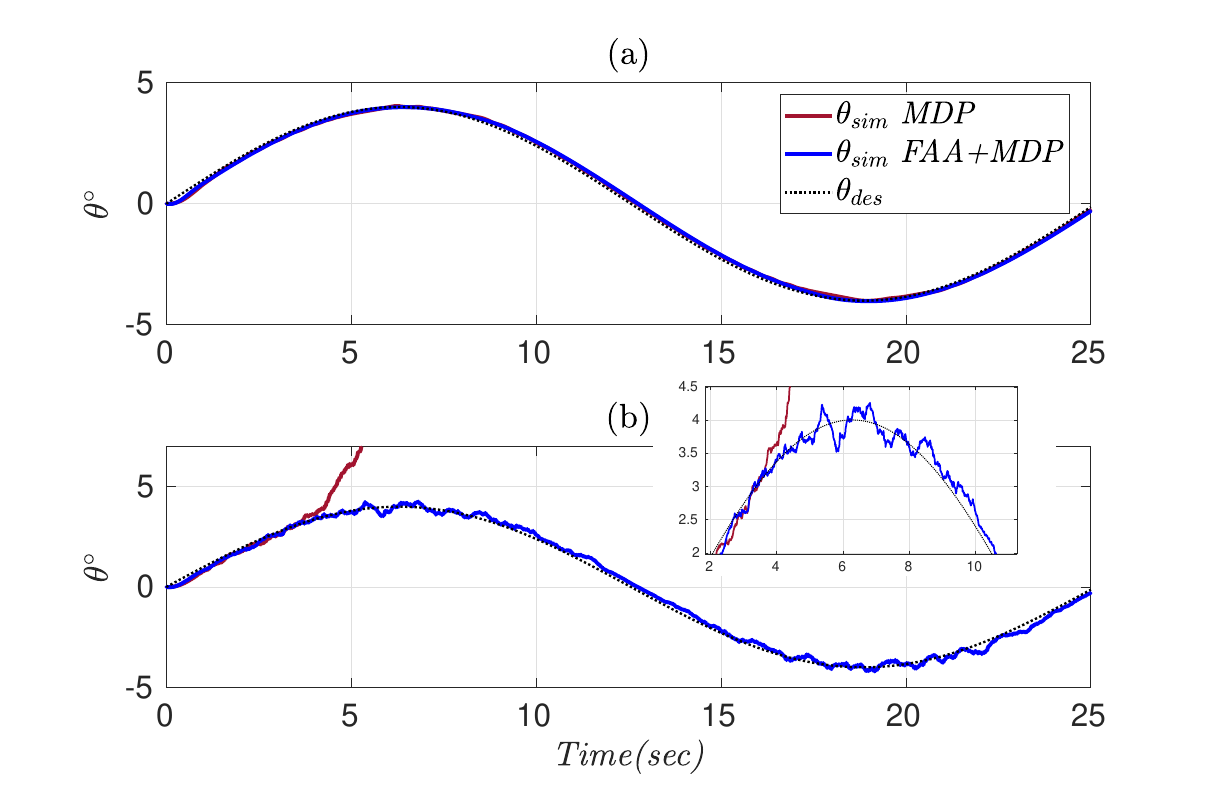}
	\caption{Variable $\theta$ tracking using MDP and FAA-improved MDP a) at normal condition, and b) in presence of sensor measurement noise and atmospheric disturbance.}
	\label{fig:7}
\end{figure}
Finally, figure (\ref{fig:7}), shows the tracking result for a variable $\theta$. The tracking of a variable pitch angle became possible by defining a virtual desired pitch angle according to the current tracking error of the system.
At first glance, it is obvious that both MDP and FAA-improved MDP were prosperous to track variable $\theta$ in normal conditions. But the elevator's high working frequency of MDP which is shown in figure (\ref{fig:7}), proves the FAA significant superiority. Apart from that, in presence of atmospheric disturbances and sensor measurement noise that means an actual flight condition, FAA-improved MDP is able to track $\theta_{des}$ and demonstrate its robustness. But in figure (\ref{fig:7}) and its zoom-in, the MDP starts to diverge when the atmospheric disturbance is applied in addition to noise.

\begin{figure}
	\includegraphics[width=9.0cm]{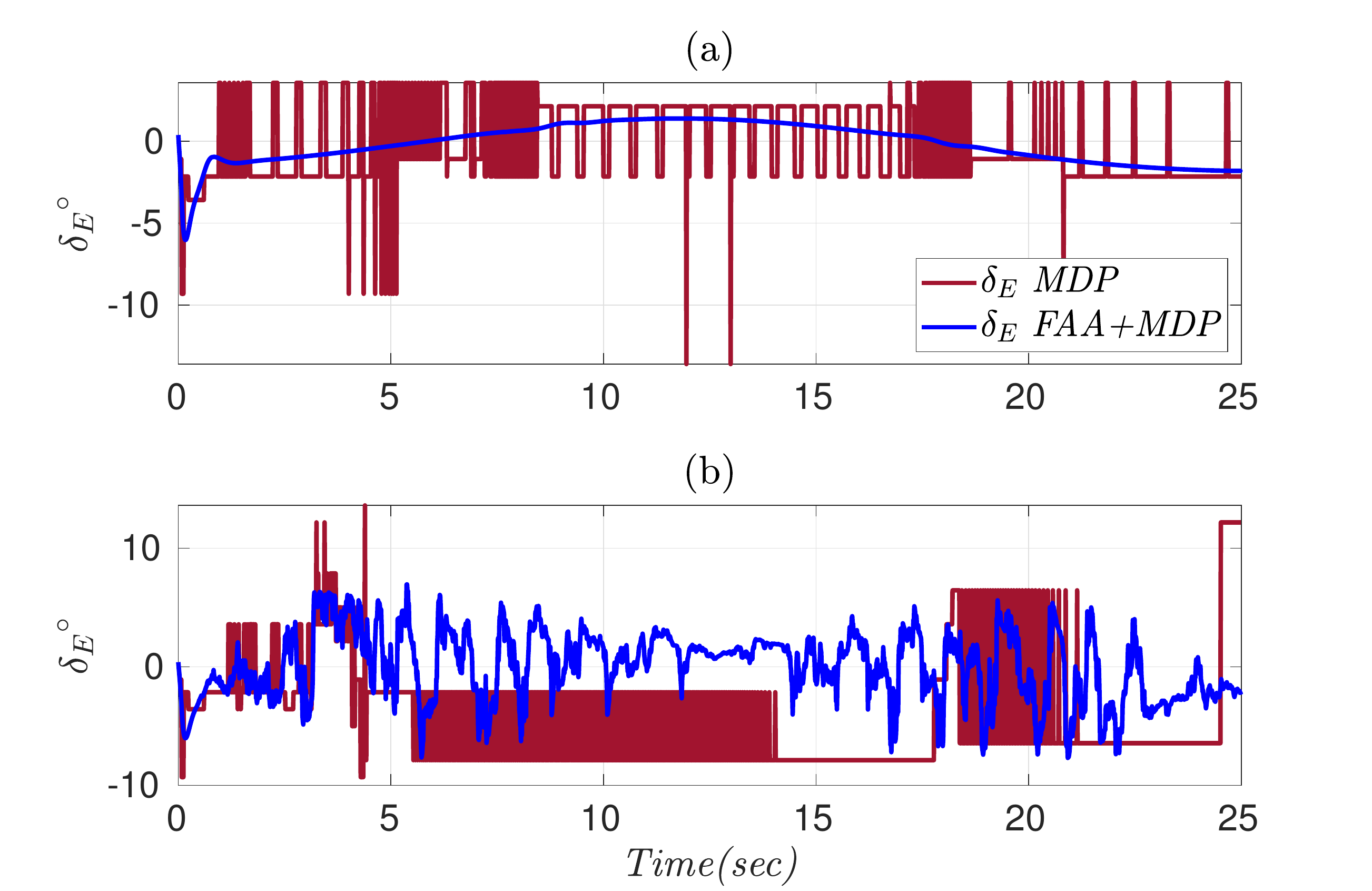}
	\caption{Elevator deflections in variable $\theta$ tracking a) at normal condition, and b) in presence of sensor measurement noise and atmospheric disturbance.}
	\label{fig:8}
\end{figure}

\section{Conclusion}
This work proposes a longitudinal attitude control method based on improved Q-learning under MDP and POMDP problem modeling. The aircraft is a novel regional Truss-braced wing airplane that is treated as an RL environment. The action selection strategy is $\epsilon$-greedy, and the reward function defines to consider $\theta$, $q$, and $\delta_{E}$. This method is verified in constant and variable $\theta_{des}$ tracking simulations where the simulation results show a comparable outcome between the FAA-improved MDP model and the PID controller. Fruitful to conclude that POMDP modeling results are poor due to imperfect problem definition. Finally, the FAA-augmented method was used to construct continuous actions to eliminate fluctuations and make robust variable angle tracking in presence of atmospheric disturbance and sensor measurement noise.

\bibliography{Paper}             







\end{document}